\documentclass{article}

\usepackage{arxiv}

\usepackage[utf8]{inputenc} 
\usepackage[T1]{fontenc}    
\usepackage{hyperref}       
\usepackage{url}            
\usepackage{booktabs}       
\usepackage{amsfonts}       
\usepackage{nicefrac}       
\usepackage{microtype}      
\usepackage{lipsum}		
\usepackage{graphicx}
\usepackage{subfigure}
\usepackage{cite}
\usepackage{amsmath,amssymb,amsthm} 
\usepackage{braket} 

\title{ The formation region of emitted $\alpha$ and heavier particles inside radioactive nuclei}

\author{ W. M. Seif $^{1,2(*)}$ and A. M. H. Abdelhady$^{1}$ \\
 $^1$Cairo University, Faculty of Science, Department of Physics, 12613 Giza, Egypt \\
 $^2$Beni-Suef University, Faculty of Navigation Science and Space Technology, \\ \, \, 62514 Beni-Suef, Egypt \\
\texttt{$^*$wseif@sci.cu.edu.eg} \\
}

\begin{document}
\maketitle

\begin{abstract}
We investigate the formation distance ($R_0$) from the center of the radioactive parent nucleus at which the emitted cluster is most probably formed. The calculations are microscopically performed starting from solving the time-independent Schr\"{o}dinger wave equation for the $\alpha$-core system, using  nuclear potential based on the Skyrme-SLy4 nucleon-nucleon interaction and folding Coulomb potential, to determine the incident and transmitted wave functions of the system. Our results advocate that the emitted cluster is mostly formed in the pre-surface region of the nucleus, under the effect of Pauli blocking from the saturated core density. The deeper $\alpha$-formation distance inside the nucleus gives rise to less preformation probability, and indicates more stable nucleus of longer half-life. Also, the $\alpha$-particle tends to be formed at a bit deeper region inside the nuclei having larger isospin asymmetry and in the closed shell nuclei. Regarding the emitted nuclei heavier than $\alpha$-particle, we find that the formation distance of the emitted clusters heavier than $\alpha$-particle increases with increasing the isospin asymmetry of the formed cluster, rather than with increasing its mass number. The partial half-life of a certain cluster-decay mode increases upon increasing either the mass number or the isospin asymmetry of the emitted cluster.
\end{abstract}


\section{Introduction}

The principle of clustering is understood as a nuclear phenomenon in states around the cluster decay thresholds and in describing the nuclear structure \cite{ref47,ref84,ref156,ref157,refw1}. A nuclear cluster might be characterized as a spatially located subsystem made out of strongly correlated nucleons, and it is described by intrinsic binding that is stronger than its external binding \cite{ref158}. It is then conceivable to think about the cluster as a solitary unit, and to depict its conduct without reference to its interior structure. One of the most dominant cluster nuclei that have been used to study the nuclear structure of heavy nuclei is the $\alpha$-cluster nucleus because of its high accuracy experimental measurements and the availability of its microscopic theory \cite{ref7,ref8,ref9}. During the $\alpha$ or heavier cluster decay process, nucleons in the parent nucleus spend part of their time as clusters, in the nuclear surface \cite{ref10,ref11}. The strongly correlated nucleons in the parent nucleus can condense to form a cluster \cite{ref12}. Thus, two neutrons and two protons can condense to form an $\alpha$ cluster, which then oscillates within the surface of the nucleus with certain quantum numbers on a shell model base. The formation probability of the condensed nucleons within the nuclear surface is determined by the preformation spectroscopic factor ($S_{\alpha(c)}$). Once formed, the light cluster will try to tunnel through the cluster-core Coulomb barrier, by knocking against it with an extremely large frequency, leaving the daughter behind. The quantum tunneling effect takes place in this stage through the penetration probability ($P$) and the assault frequency ($\nu$). Both the preformation and the penetration probabilities can be numerically obtained from the incident and transmitted wave functions of the system, which are obtained from the numerical solution of time-independent Schr\"{o}dinger equation for a certain $\alpha$- or heavier cluster-decay. The assault frequency is defined as the inverse of the time taken for the formed cluster to make one oscillation within the internal pocket region of the potential. Thereafter, the half-life can be estimated in terms of the preformation probability, the assault frequency, and penetration probability.

Commonly, the probability of forming the emitted particle from strongly correlated surface nucleons in the parent nucleus is called the preformation probability or the preformation factor. In the literature it may also be called the cluster-formation probability, the spectroscopic factor, or the amount of clustering. The cluster preformation probability is very important quantity because it reflects information about the nuclear structure. For instance, it is a good indicator of the deformation in the nuclei participating in the decay process. The preformation probability increases if the cluster is formed from the nucleons belonging to the last open shells, which lead to deform the nuclear surface \cite{ref_A1}. In addition to its model dependence, the uncertainty in the estimated preformation probability increases when the involved nuclei are deformed in their ground states or have larger isospin asymmetry \cite{ref_A2,ref_A3}. Microscopically, the preformation probability can be calculated in terms of the formation amplitude of the amount of clustering of the emitted and daughter nuclei as two distinguishable entities inside the parent nucleus \cite{ref_A6}. The formation amplitude represents the projection of the parent wave function with respect to the anti-symmetrized product of the identified wave functions of the cluster and core fragments. Using the R-matrix description \cite{ref_A7}, the formation amplitude of the $\alpha$-cluster in $^{212}$Po has been calculated using high configuration mixing of harmonic-oscillator bases \cite{ref_A10}. The calculations of the formation amplitude were improved using the multistep shell model method with including pairing and mutual interactions between the valence nucleons \cite{ref_A6,ref_A7}, as well as considering the high-lying states. The dimensions of matrix elements were reduced using surface delta interaction in truncated model space \cite{ref_A6}. The existence of $\alpha$-core structure has been confirmed in the frame work of both the shell model and the cluster model within bases of large dimensions, and using Gaussain-bases and large number of configurations \cite{ref_A8,ref_A9}. Moreover, the $\alpha$ clustering has been described in many cluster-formation states \cite{refA27}. 

The emitted light nucleus is most probably clustered at appropriately far distance from the center of the core nucleus, near the surface of the parent nucleus, due to Pauli blocking impacts from the core density \cite{ref10,ref12,ref123,ch1_ref165}. In the present work, we will investigate the location of forming the emitted cluster inside the parent nucleus. We shall use the relative motion  wave function of the cluster-core system that obtained from the time-independent Schr\"{o}dinger wave equation to estimate the distance, from the center of the parent nucleus, at which the cluster is formed with a reasonable preformation probability.

\section{Theoretical Formalism}

The interaction potential between the emitted cluster and daughter nucleus is a basic ingredient to study the decay process of a certain nucleus. After constructing the cluster-core interaction potential, it can be implemented in the time-independent Schr\"{o}dinger wave equation to determine the incident and transmitted wave functions of the cluster-core system. Based on the Skyrme energy density functionals, the nuclear interaction potential, as a function of the separation distance $r$(fm) between the centers of mass of the interacting nuclei, is obtained by the difference between the energy expectation value $E$ of the composite system at a finite separation distance $r$ and that of individual separated nuclei at $r = \infty$ \cite{refA27,refA29,refA31},
\begin{align}\label{eq1}
V_N(r) &= E(r) - E(\infty) \nonumber \\
&= \int \left\lbrace  \mathcal{H}\left[ \rho_{pc}(\vec{x}) + \rho_{pD}(r, \vec{x}), \, \rho_{nc}(\vec{x}) + \rho_{nD}(r, \vec{x}) \right]  \right. \nonumber \\ & \qquad \left. -  \mathcal{H}_c\left[ \rho_{pc}(\vec{x}) , \, \rho_{nc}( \vec{x}) \right] - \mathcal{H}_D\left[ \rho_{pD}(\vec{x}) , \, \rho_{nD}( \vec{x}) \right] \right\rbrace d\vec{x}.
\end{align}
$\mathcal{H}$, $\mathcal{H}_c$ and $\mathcal{H}_D$ in Eq.(\ref{eq1}) define the energy density functionals of the composite system, the formed cluster, and the daughter nucleus, respectively. $\rho_{ij}(i=p,n;j=c=D)$ represent the proton ($p$) and neutron ($n$) density distributions of both the emitted cluster ($c$) and the daughter nucleus ($D$). The Skyrme energy functional includes the kinetic and the nuclear (Sky) contributions, 
\begin{align}\label{eq2}
\mathcal{H} \left( \rho_i, \tau_i, \vec{J}_i\right) = \frac{\hbar^2}{2m}\sum_{i=n,p}\tau_i\left(\rho_i,\vec{\nabla}\rho_i, \nabla^2 \rho_i \right) +  \mathcal{H}_{Sky}\left( \rho_i, \tau_i, \vec{J}_i\right) + \mathcal{H}_{C}^{exch}\left( \rho_p\right). 
\end{align}
Here, $\tau_i$ and $\vec{J}_i$ respectively define the kinetic energy and the spin-orbit densities \cite{refA27,refA30,refA32}. Regarding nuclear part of the energy-density functional, we shall use the Skyrme-SLy4 parameterization \cite{refA33} of the effective nucleon-nucleon, which includes zero- and finite-range, density-dependent, effective-mass, spin-orbit, tensor, and surface gradient terms. The last term in Eq.(\ref{eq2}) considers the exchange Coulomb energy \cite{refA34,ref_A1}. The direct part of the Coulomb potential can be obtained by folding the proton-proton Coulomb interaction through the proton density distributions of the interacting nuclei \cite{ref_A2}, 
\begin{align}\label{eq3}
V_c(r) =\int d\vec{r}_1 \int d\vec{r}_2 \frac{e^2}{|\vec{r} + \vec{r}_2 + \vec{r}_1|} \, \rho_{pc}(\vec{r}_1) \, \rho_{pc}(\vec{r}_2).
\end{align}
More details concerning the method of calculating the Coulomb and nuclear parts of the interaction potential can be found in Refs. \cite{ref_A3,refA27,refA26}. The neutron (proton) density distribution of the involved nuclei heavier than $\alpha$-particle can be expressed in the two-parameter Fermi form, 
\begin{align}
\rho_{n(p)} (r) = \rho_{0n(p)} \left[ 1+ e^{\frac{\left(r-R_{n(p)}\right)}{a_{n(p)}}}\right]^{-1}.
\end{align}
Based on a fit to a huge number of nuclear density distributions, which are obtained using Hartree-Fock calculations in terms of the Skyrme-SLy4 \textit{NN} interaction, the half-density radii ($R_{n(p)}$) and diffuseness $a_{n(p)}$ of finite nuclei have been parameterized as \cite{ref_A4},
\begin{align}\label{eq5}
& R_n (\text{fm}) = 0.953\, N^{(1/3)} + 0.015\, Z + 0.774, \nonumber \\
& R_p (\text{fm}) = 1.322\, Z^{(1/3)} + 0.007\, N + 0.022,  \\
& a_n (\text{fm}) = 0.446 + 0.072 \left(\frac{N}{Z}\right), \nonumber \\
& a_p (\text{fm}) = 0.449 + 0.071 \left(\frac{Z}{N}\right). \nonumber
\end{align}
The saturation density $\rho_{0n(p)}$ is evaluated through normalizing the density to the corresponding nucleon number. This parameterization takes advantage of considering the isospin asymmetry dependence of the nuclear density distributions by giving the proton (neutron) density distribution as a function of both $Z$ and $N$ together. The $\alpha$-particle density is usually taken as Gaussian distribution that is parameterized via electron scattering data \cite{ref_A5}. For the favored decay modes with no angular momentum transferred by the emitted light cluster, we add the nuclear and Coulomb potentials together to construct the total potential,
\begin{align}\label{eq6}
V(r) = V_N(r) + V_C(r).
\end{align}
The total potential $V_T(r)$ is characterized by three classical turning points, $r_{i=1,2,3}$(fm), at which $V_T(r_i)$ equals the $Q$-value ($Q_C$) of the decay process. Once we construct the total interaction potential, we implement it in the radial Schr\"{o}dinger wave equation for the cluster-core dinuclear system,  
\begin{align}\label{eq7}
-\frac{\hbar^2}{2\mu}\frac{d^2}{dr^2} u_l(r) \,+\, \left( V(r) + \frac{l(l+1) \hbar^2}{2\mu r^2} \right) u_{l}(r) = E \, u_{l}(r).
\end{align}
Here, $u_l(r)$ represents the radial wave function, $\psi = Y_{lm}(\theta,\phi) u_l(r)/r$, that can be obtained by numerically solving Eq.(\ref{eq7}). After determining the incident ($u_{li}$) and the transmitted ($u_{lt}$) wave functions at both sides of the Coulomb barrier, we can calculate the penetration probability in terms of their squared amplitudes \cite{ref_A6,ref_A7}, 
\begin{align}\label{eq8}
P = \frac{|A(u_{lt}(r_3))|^2}{|A(u_{li}(r_2))|^2}.
\end{align}
The assault frequency with which the formed cluster hits the Coulomb barrier can be obtained, using the Wentzel-Kramers-Brillouin (WKB) approximation. as the inverse of the time taken by the cluster to make one oscillation within the internal pocket of the interaction potential, between $r_1$ and $r_2$ \cite{ref_A8,ref_A9},
\begin{align}\label{eq9}
\nu = \left[ \int_{r_1}^{r_2} \frac{2\mu}{\hbar k(r)} dr \right]^{-1}, 
\end{align}
The wave number $k$ in Eq.(\ref{eq9}) is determined as $k(r) = \sqrt{ 2 \mu |V(r)-Q_C|/ \hbar^2}$. 
The half-life of the nucleus can be obtained in terms of $P$, $\nu$, and the preformation probability ($S$) of the formed cluster as, 
\begin{align}\label{eq10}
T_{1/2} = \frac{\ln 2}{S\, \nu \, P}.
\end{align}
The preformation factor can be estimated by integrating the squared incident wave function $|u_{li}|^2$ from the origin to a certain distance $R_0$, from the parent nucleus center, at which the cluster is formed \cite{ref_A6}. To find this distance, we can employ an iterative procedure in terms of the experimentally observed half-life and the calculated penetration probability and assault frequency through the relation,
\begin{align}\label{eq11}
\int\limits_{0}^{R_0} |u_l(r)|^2 \, dr = \frac{\ln(2)}{P \, \nu \, T_{1/2}^{exp}},
\end{align}
The left-hand side of Eq.(\ref{eq11}) represents the cluster preformation probability at the distance $R_0$, from the center of the parent nucleus.
\section{Results and discussions}

Within the preformed cluster model, we try to estimate the most probable distance ($R_0$), from the center of the parent nucleus, at which the emitted light nucleus is formed. Towards this goal, we use the Skyrme-SLy4 nucleon-nucleon interaction to construct the total interaction potential for cluster-daughter system, in the framework of the Hamiltonian energy density formalism for the nuclear part of the potential and the folding procedure for its Coulomb part. We implement the total interaction potential into the relative motion Schr\"{o}dinger wave equation accompanied with the Bohr-Sommerfeld quantization condition. We numerically solve the time-independent Schrodinger wave equation of the decaying system, to extract the radial wave function ($u_l(r)$) of the quasi-bound state. The formation distance, from the parent nucleus center, will be iteratively estimated using Eq.(\ref{eq11}). 
\begin{figure}
\centering
\includegraphics[width=14cm]{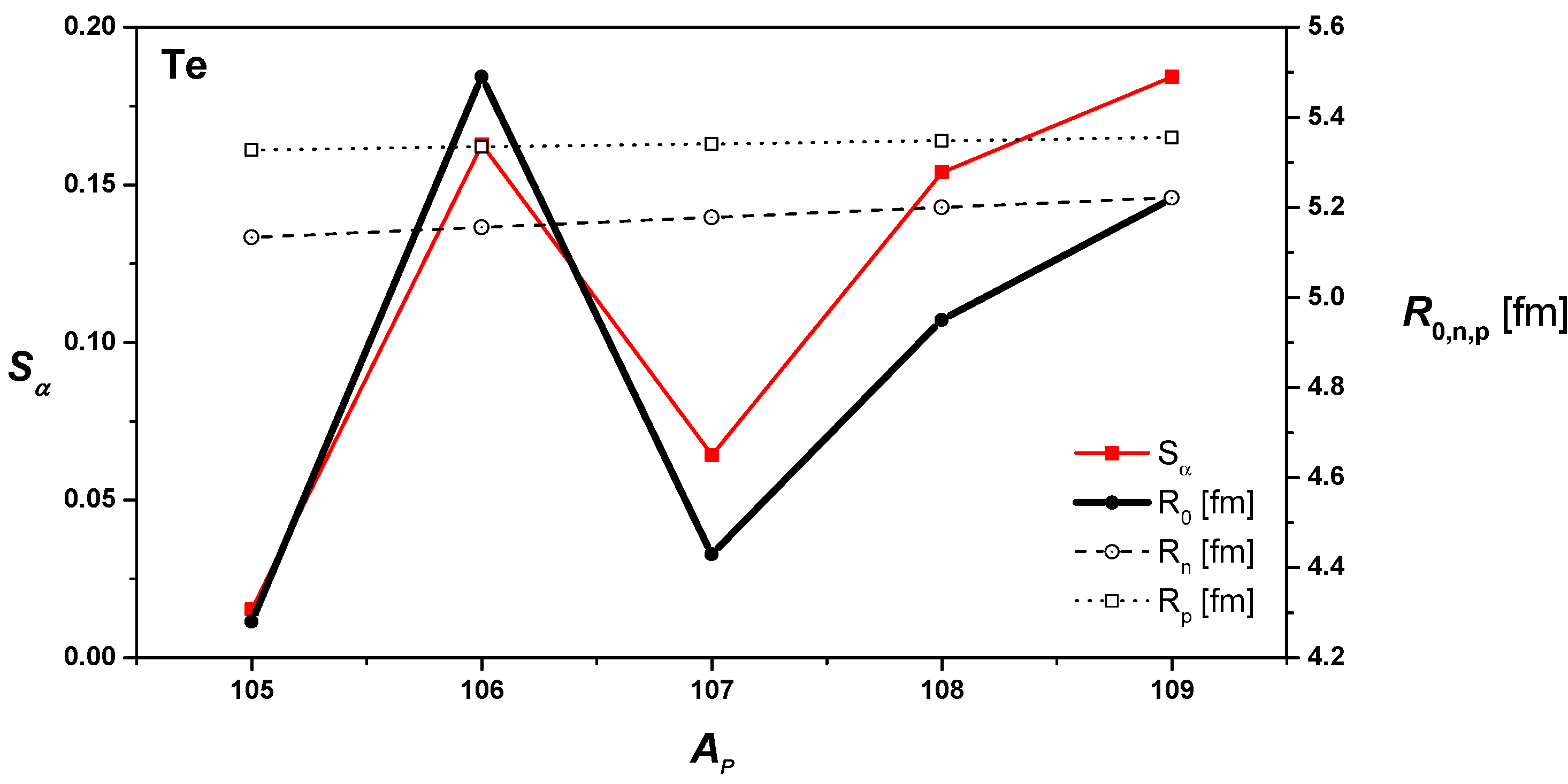}
\caption{The variation of the indicated $\alpha$-formation distance $R_0$ (Eq.(\ref{eq11})) from the center of the parent nucleus for the favorite g.s. to g.s. $\alpha$-decay modes of Tellurium isotopes, the proton ($R_p$) and neutron ($R_n$) half-density radii (Eq.(\ref{eq5})) of parent Te isotopes, and the evaluated values of the $\alpha$-preformation probability $S_{\alpha}$ at the indicated $R_0$ (Eq.(\ref{eq11})), against the parent mass number.}
\label{fig1}
\end{figure}

We start our investigation of the estimated formation distance by the preformation of the $\alpha$-particle inside the $^{105-109}$Te isotopes, which exhibit ground state (g.s.) to ground state favorite $\alpha$-decay modes. Here, the produced daughter isotopes ($^{101-105}$Sn) are around double magic number and have close numbers of neutrons and protons, which occupy similar energy levels. Fig.\ref{fig1} shows the variation of the formation distance $R_0$ from the centers of the Tellurium isotopes, as estimated using Eq.(\ref{eq11}), against the parent mass number.  In the same figure, we display the corresponding evaluated values of the $\alpha$-preformation probability at the estimated distances, Eq.(\ref{eq11}). The estimated preformation probability in Fig.\ref{fig1} coincides with that obtained in different previous studies \cite{ref7,ch1_ref148,ch1_ref149}. Also the proton ($R_p$) and neutron ($R_p$) half-density radii of the parent nuclei, which are given by Eq.(\ref{eq5}), are shown in Fig.\ref{fig1}. Figure \ref{fig1} shows that the $\alpha$-particle is formed within the region around the proton and neutron half-density radii of the parent nucleus. While the average proton and neutron half-density radii of the investigated Te isotopes are 5.34 fm and 5.18 fm, respectively, the average formation distance inside them is about 4.87 fm, which is smaller than the average $R_p$ and $R_n$ by about 0.47 fm and 0.3 fm. We find that the behavior of the calculated preformation probability with $A_P$ resembles that of the estimated formation distance. The preformation probability increases when the $\alpha$-particle is indicated to be formed at larger distance from the center of the parent nucleus, within its surface region. The smaller preformation probability refers to shorter indicated formation distance. We recall here that we are talking about different $\alpha$-daughter systems not only with different wave functions.
\begin{figure}
    \centering
    \subfigure[\label{fig2a}]{ \includegraphics[width=7.6cm]{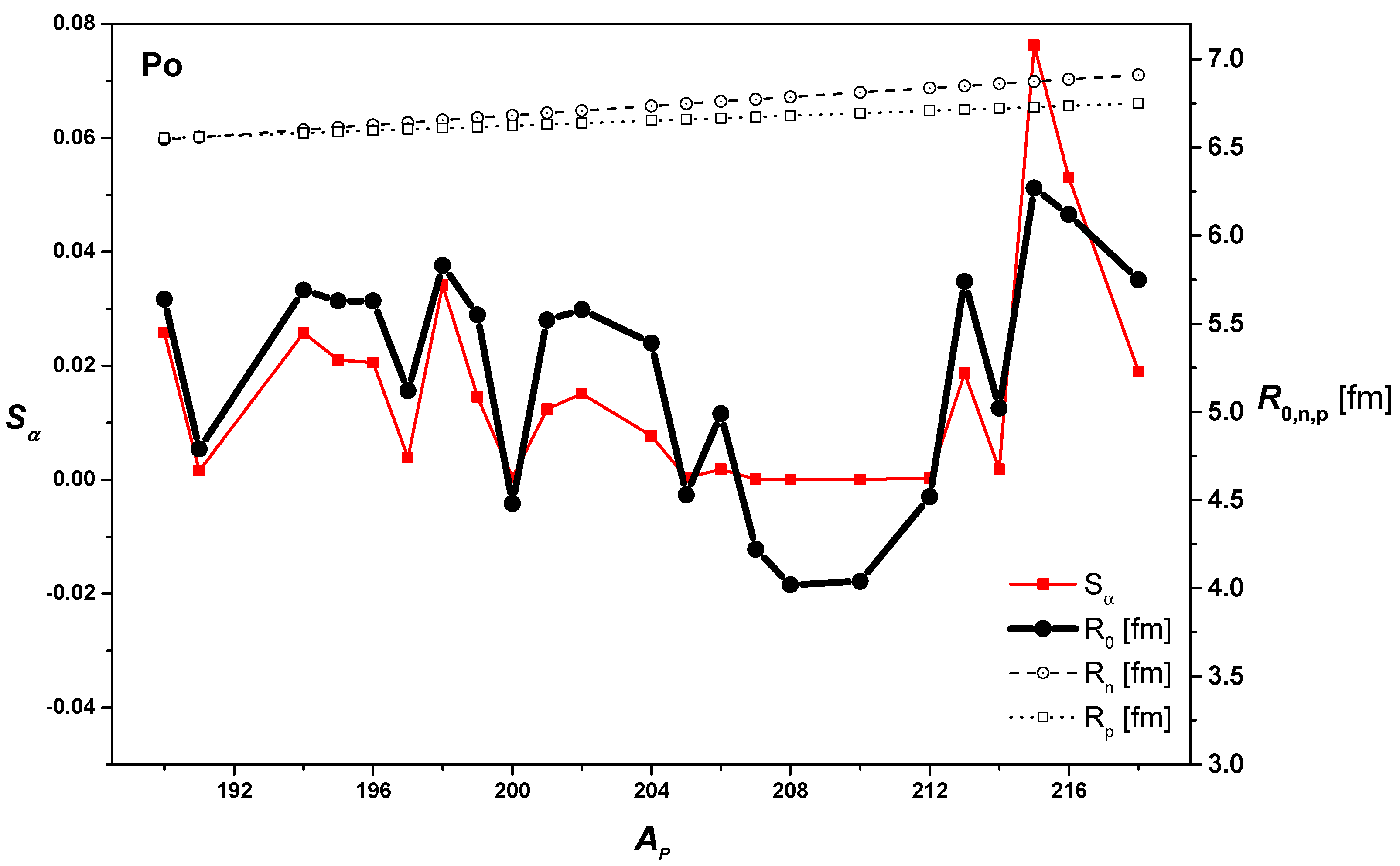} }%
    \subfigure[\label{fig2b}]{  \includegraphics[width=7.6cm]{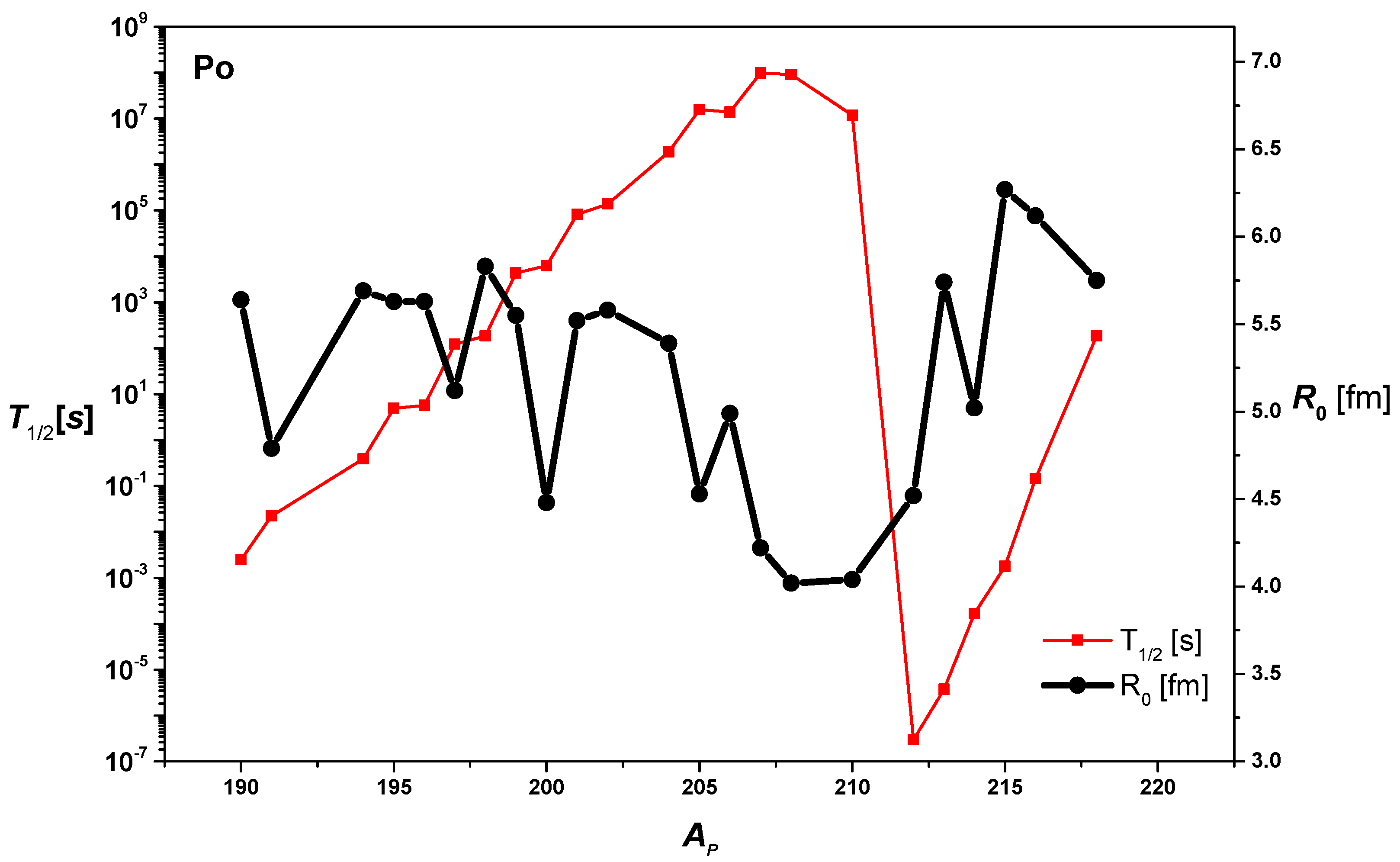} }%
    \caption{(a) Same as in Fig.\ref{fig1} but for the favorite g.s. to g.s. $\alpha$-decay modes of Polonium isotopes. (b) The variation of the $\alpha$-decay half-lives of the Po isotopes and the corresponding formation distances, against the parent mass number. }%
    \label{fig2}%
\end{figure}
 
Now we come to the favorite g.s. to g.s. decays of the even-even and even-odd $^{186,190,191,194-202,204-208,210,212-216,218,219}$Po nuclei, which are heavier and have larger isospin asymmetry than the Te isotopes. We show in Fig.\ref{fig2a} the variation of the estimated formation distance, the $\alpha$-preformation probability, and the proton and neutron half-density radii of the parent nuclei, as functions of the parent mass number. The estimated values of the preformation probability in Fig.\ref{fig2a} based on the numerical solution of the time-independent Schr\"{o}dinger equation of the $\alpha$+Pb systems agree with those obtained in previous studies \cite{ch1_ref169,ch1_ref170,ch1_ref171,ch1_ref172,refA27}. While the average proton and neutron half-density radii of the investigated Polonium isotopes in Fig.\ref{fig2a} are 6.65 fm and 6.73 fm, respectively, the average formation distance from their centers is about 5.22 fm, which is smaller than the average $R_p$ and $R_n$ by about 1.43 fm and 1.51 fm. This means that the $\alpha$-particle is formed at a bit deeper region inside the Po isotopes relative to the formation distance inside the Te isotopes. This is due to that the Po isotopes are of larger isospin asymmetry than the Te isotopes. Again, we see that the behavior of the preformation probability with $A_P$ simulates that of the formation distance, where the shorter formation distance reflects less preformation probability. The formation region is deeper inside the Po isotopes of neutron numbers near the 126-neutron closed shell. Figure \ref{fig2b} shows the variation of the $\alpha$-decay half-lives of the Po isotopes and the corresponding formation distances inside them. As shown in Fig.\ref{fig2a}, the behavior of the formation distance with $A_P$ inversely reflex that of the half-life, where the shorter formation distance indicates more stable nucleus.
 
\begin{figure}
\centering
\qquad \qquad \includegraphics[width=14cm]{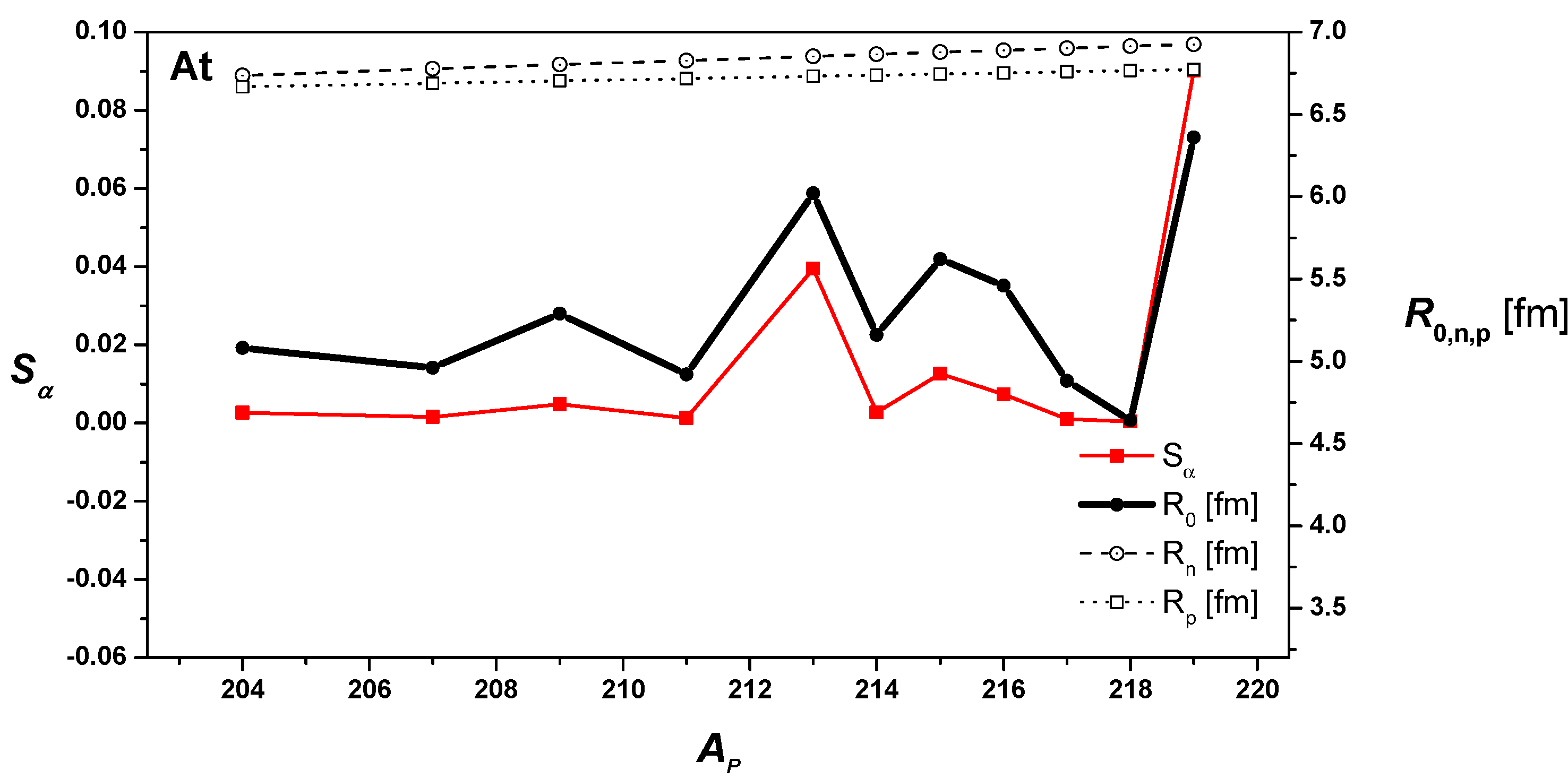}
\caption{ (a) Same as in Fig.\ref{fig1} but for the favorite g.s. to g.s. $\alpha$-decay modes of Astatine isotopes. } 
\label{fig3}
\end{figure}
 
The estimated formation distance and the calculated preformation probability for the favorite g.s. to g.s. decays odd-even and odd-odd $^{204,207,209,211,213-219}$At isotopes are displayed in Fig.\ref{fig3}. These Astatine isotopes have isospin asymmetry comparable to (larger than) that of the Polonium (Tellurium) isotopes. The estimated values of $S_{\alpha}$ in Fig.\ref{fig3} come to an agreement with those yielded in other studies \cite{ch1_ref173,ch1_ref172,ch1_ref152}. The average proton and neutron half-density radii of the Astatine isotopes presented in Fig.\ref{fig3} are 6.73 fm and 6.85 fm, respectively. The average formation distance from their centers is about 5.31 fm. The average formation distance is then smaller than the average $R_p$ and $R_n$ by about 1.42 fm and 1.54 fm. This shows that the emitted $\alpha$-particle is formed within the same pre-surface region pointed out for the Polonium isotopes.

\begin{figure}
     \centering
     \subfigure[\label{fig4a}]{ \includegraphics[width=8.15cm]{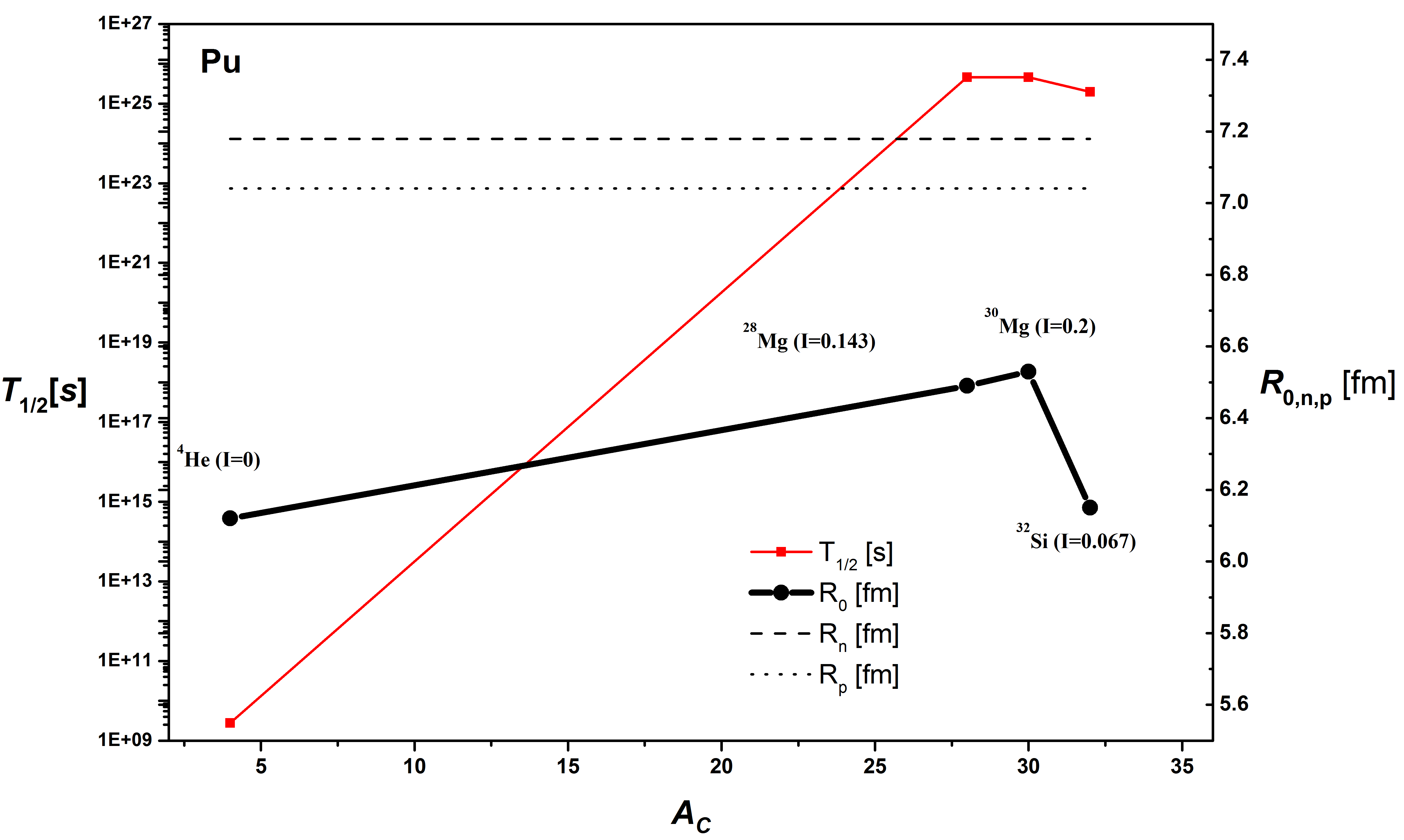} }%
     \subfigure[\label{fig4b}]{ \includegraphics[width=6.95cm]{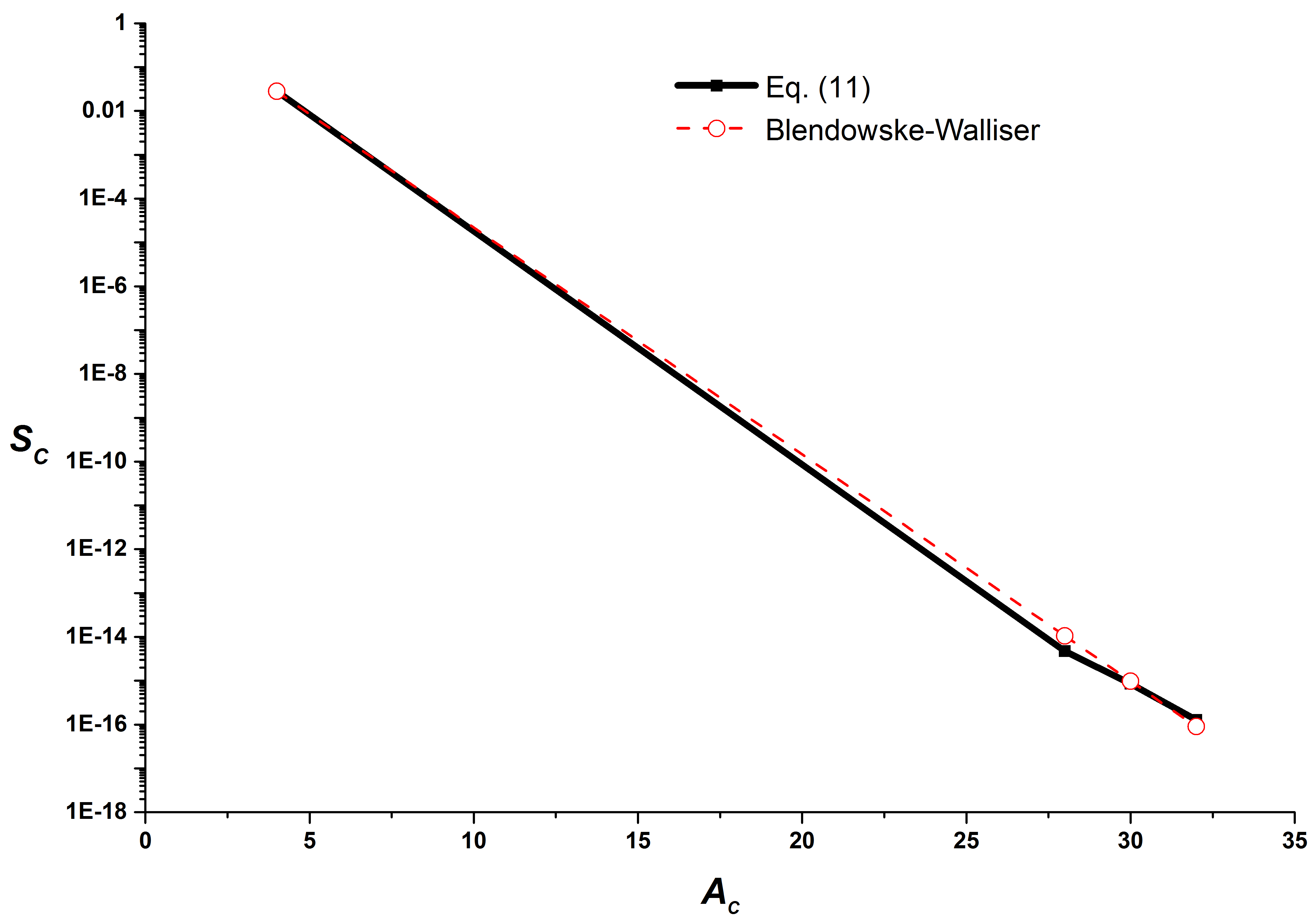} }%
     \caption{(a) The variation of the indicated formation distance $R_0$ from the center of the $^{238}$Pu parent nucleus in its $\alpha$, $^{28,30}$Mg and $^{32}$Si  decay modes, the proton  and neutron half-density radii of $^{238}$Pu, and the corresponding observed partial half-lives, against the mass number of the emitted cluster. (b) The microscopically estimated cluster-preformation probability $S_C$ in $^{238}$Pu.  The red dashed line shows the Blendowske-Walliser behavior of $S_C$ ($^{238}$Pu) in terms of the microscopically estimated $S_{\alpha}$. }%
     \label{fig4}%
 \end{figure}

We turn now to the emitted nuclei heavier than $\alpha$-particle. One of the heavy nuclei that are observed to emit heavy clusters is the $^{238}$Pu nucleus, which emits $^{28,30}$Mg and $^{32}$Si nuclei with branching ratios of about 6 x 10$^{-15}$ \% and 1.4 x 10$^{-14}$ \%, respectively, in addition to its principal $\alpha$-decay mode ($\approx$ 100\%) \cite{ch1_ref87}. In Fig.\ref{fig4a} we show the estimated formation distance and the observed half-life as functions of the mass number of formed cluster ($A_C$). Shown in Fig.\ref{fig4b} are the estimated values of the  preformation probability based on the numerical solution of the time-independent Schr\"{o}dinger equation of the $\alpha$ + $^{234}$U ($Q_{\alpha}$ = 5.593 MeV), $^{28}$Mg + $^{210}$Pb ($Q_{(^{28}\text{Mg})}$ = 75.912 MeV), $^{30}$Mg + $^{208}$Pb ($Q_{(^{30}\text{Mg})}$ = 76.797 MeV), and $^{32}$Si + $^{206}$Hg ($Q_{(^{32}\text{Si})}$ = 91.188 MeV) systems. Figure \ref{fig4a} shows that the formation distance inside the nucleus does not vary that much with increasing the mass number of the formed cluster. However, the formation distance of the heavy clusters increases with increasing the isospin asymmetry of the formed cluster. For instance, the $\alpha$ and $^{28}$Si clusters, which have isospin asymmetry parameter $I=(N-Z)/A=0$ and $I=0.067$, respectively, are estimated to be formed at the short distance of 6.12 fm and 6.15 fm, from the center of the $^{238}$Pu nucleus. On the other hand, the $^{28}$Mg($I=0.143$)  and $^{30}$Mg($I=0.2$) clusters are estimated to be formed at larger distances of 6.49 fm and 6.53 fm, respectively.  We recall here that the local isospin asymmetry increases on the surface and tail regions of the nucleus. Then, the binding energy of the surface and tail nucleons become less than that of those in the interior region, where the energy per nucleon decreases with increasing the isospin asymmetry \cite{refA27}. This, in addition to the Pauli blocking effects from the saturated core density \cite{ref10,ref12,ref123}, endorses the formation of the cluster near the surface nuclear region. In addition to its $A_C$ dependence, the observed partial half-lives of $^{238}$Pu against its cluster-decay modes also increase with increasing the isospin asymmetry of the emitted cluster, as shown in Fig.\ref{fig4a}. Figure \ref{fig4b} shows the cluster preformation probability that is microscopically estimated based on Eq.(\ref{eq11}) and that calculated using the phenomenological formula of Blendowske and Walliser (BW), $S_C= (S_{\alpha})^{\frac{A_C-1}{3}}$ \cite{ch1_ref52,ch1_ref177}, in terms of the microscopically estimated $\alpha$ preformation probability. As seen in Fig.\ref{fig4b}, the microscopically obtained preformation probability approves its $A_C$ dependence given by the BW relation.

\section{Conclusion}

Using the wave functions obtained by the numerical solution of the stationary Schr\"{o}dinger wave equation for the decaying system, we have investigated the formation distance from the center of the parent nucleus at which the emitted cluster is formed. The results endorse that the emitted cluster is often formed in the pre-surface region of the parent nucleus under the effect of Pauli blocking from the saturated core density. While the estimated preformation probability of the $\alpha$-particle mimics the behavior of its formation distance with the mass number of the parent nucleus, the observed half-life inversely reflex this behavior. Thus, the deeper $\alpha$-formation distance inside the parent nucleus leads to less preformation probability and longer partial half-life. Further, the $\alpha$-cluster tends to be formed at a bit deeper sub-saturation density region in the nuclei of larger isospin asymmetry and in the closed shell nuclei. On the other hand, we found that the cluster-formation distance increases with increasing the isospin asymmetry of the formed cluster, but not with increasing the cluster mass number. However, for a certain cluster-decay mode, the partial half-life increases with both the mass number of the emitted cluster and its isospin asymmetry.
\newpage
\bibliographystyle{unsrt}  


\end{document}